 \documentclass [12pt,a4paper      ]{article}
\usepackage{times}

\DeclareFontFamily{OT1}{times}{}
\DeclareFontShape {OT1}{times}{m }{n }{ <-> ptmr }{}
\DeclareFontShape {OT1}{times}{bx}{n }{ <-> ptmb }{}
\DeclareFontShape {OT1}{times}{m }{it}{ <-> ptmri}{}
\DeclareFontShape {OT1}{times}{bx}{it}{ <-> ptmbi}{}
\usepackage{amsmath}
\usepackage{amsfonts}
\usepackage{amssymb}
\usepackage{latexsym}
\usepackage{isridef} 
\numberwithin{equation}{section}               

\begin{document}

\title{\bf\vspace{-2.5cm} A possible fundamental explanation
                          of electroweak unification }

\author{
         {\bf Andre Gsponer}\\
         {\it Independent Scientific Research Institute}\\ 
         {\it Oxford, OX4 4YS, England}
       }

\date{ISRI-08-07.10 ~~ November 24, 2008}

\maketitle

\begin{abstract}

Electroweak unification is implied by the local structure theorem of distribution theory applied to the causal interval $R=X-Z$ between two space-time points $X$ and $Z$.  Taking $R$ as generating function, the potentials of the electromagnetic and weak interactions are obtained by applying the causal d'Alembertian operator on $R$, and both potentials can be given a proper physical interpretation provided all calculations are made in the context of nonlinear generalized functions.

\end{abstract}

\section{Introduction}
\label{int:0}

In the current `standard model' of elementary particles, electroweak unification does not derive from a fundamental principle.  Rather, it is implemented by construction.  In this paper we show that electroweak unification is implied by a physically simple and natural principle supported by the theory of distribution and its pre-eminent generalization, Colombeau's theory of nonlinear generalized functions \cite{COLOM1990-,GROSS2001-,GSPON2006B}:
\begin{quote}
{\bf Principle [Electroweak unification]} \emph{
Let $e$ be the electric charge and let $\Phi(X,Z)$ be $e/2$ times the causal four-interval $R=X-Z$ between an observation point $X$ and an arbitrarily moving point $Z(\tau)$, i.e., such that $R\CON{R}=0$.  Let $\square$ be the causal d'Alembertian operator.  Then
\begin{equation} \label{int:1}
  \square  \Phi(X,Z) = \phi^{\text{\emph{LW}}}(X,Z) +  \phi^{\text{\emph{WI}}}(X,Z),
\end{equation}
where $\phi^{\text{\emph{LW}}}$ is the Li\'enard-Wiechert potential of an arbitrarily moving point-charge, and $\phi^{\text{\emph{WI}}}$ a maximally parity violating potential which, interpreted as a nonlinear generalized function, corresponds to weak interactions in the Fermi limit, provided the equations of motions of $Z(\tau)$ are $SU(2)$ gauge-invariant.  
Mathematically, Eq.\,\eqref{int:1} means that $\phi^{\text{\emph{LW}}}$ and $\phi^{\text{\emph{WI}}}$ are nonlinear generalized functions generated by means of the partial differential operator $\square$ from the generating function $\Phi(X,Z)$, which is a continuous function in the space-time variables $X,Z \in \mathbb{R}^4$.  Thus, electromagnetic and weak interaction cannot exist independently from each other, and therefore derive from a common seed, the  causal four-interval $R=X-Z$.
}
\end{quote}

In order to vindicate this principle we shall need to recall one of the most basic theorems of distribution theory: Laurent Schwartz's `local structure theorem,' which states that \emph{``any distribution is locally a partial derivative of a continuous function''} \cite[Theorems XXI and  XXVI]{SCHWA1966-}.  Courant and Hilbert call such a continuous function a `generating function' \cite[p.\,781]{COURA1962-}, and Schwartz's theorem the `main representation theorem' \cite[p.\,799]{COURA1962-}.  This theorem will be applied to the Coulomb and Lienard-Wiechert potentials in Sec.\,\ref{gen:0}.

In Sec.\,\ref{def:0} we give some definitions and introduce an explicit biquaternionic representation of null-intervals which will facilitate the physical interpretation of our results.

In Sec.\,\ref{cau:0} the d'Alembertian of the null-interval $R$ is calculated, and the reason why the nonlinear generalized function setting is required to interpret the results is explained.

In Sec.\,\ref{ele:0} the non-electromagnetic part of the potential deriving from $R$ is shown to correspond to weak interactions in the Fermi limit.

Finally, in Sec.\,\ref{dis:0} we discuss some of the shortcomings of the developements presented in the paper, and what should be done in order to get a complete theory of electroweak interactions.

\section{Generating functions of the Coulomb potential}
\label{gen:0}

Let the Coulomb potential $e/r$ be interpreted as a distribution $\phi(r)$, where $e$ is the electric charge of an electron at rest at the origin of a polar coordinate system and $r = |\vec{r}\,|$ the modulus of the radius vector.  Then there must be a generating function $\Phi(r) \in \mathcal{C}^0$, not necessarily unique, of which $\phi(r)$ is a partial derivative.  For example, if  $\HU$ is the Heaviside function and $e/r$ is defined as the weak limit of the sequence of distributions \cite[p.\,144]{TEMPL1953-},
\begin{equation} \label{gen:1}
  \phi(r) \DEF \frac{e}{r} \lim_{a \rightarrow 0} \HU(r-a),
\end{equation}
which is the most commonly used distributional representation of the Coulomb potential, then $\phi(r)$ is the distributional-derivative of
\begin{equation} \label{gen:2}
  \Phi(r) = e\lim_{a \rightarrow 0} \log(r/a) \HU(r-a),
\end{equation}
a $\mathcal{C}^0(\mathbb{R})$ function $\forall r \geq 0$.  Indeed, calculating its $r$-derivative we get
\begin{equation} \label{gen:3}
  \frac{\partial}{\partial r} \Phi(r) = e\lim_{a \rightarrow 0}
     \Bigl( \frac{e}{r} \HU(r-a)  +  \log(r/a) \delta(r-a) \Bigr),
\end{equation}
where the second term yields zero when evaluated against a test function in $\mathbb{R}^3$. Equation \eqref{gen:3} is thus distributionally equivalent to Eq.\,\eqref{gen:1}.


   Distributions being equivalence classes, there are of course many other possible representations of $\phi$.  To simplify calculations, we will use in this paper representations based on the generalized function $\UPS$ introduced in Ref.\,\cite{GSPON2006B,GSPON2008B}, i.e.,\footnote{Intuitively, $\Upsilon(r)$ can be seen as equivalent to the sign function ${\rm sgn}(r)$ for $r \geq 0$.}
\begin{eqnarray} \label{gen:4}
   \UPS(r) \DEF 
         \begin{cases}
         \text{undefined}   &   r < 0,\\
                      0     &   r = 0,\\
                     +1     &   r > 0,
         \end{cases}
   \qquad\text{and}\qquad
   \UPS'(r) \DEF 
   \frac{d}{dr}\Upsilon(r) = \delta(r),
\end{eqnarray}
which explicitly specifies how to consistently differentiate at $r=0$.  Thus, instead of  (\ref{gen:1}--\ref{gen:2}), we may take as representatives
\begin{equation} \label{gen:5}
  \phi(r) \DEF \frac{e}{r} \UPS(r),
  \qquad \text{and} \qquad
  \Phi(r) = \frac{e}{r} \log(r/a) \UPS(r).
\end{equation}
These expressions can be written in invariant form by introducing the four-velocity $\dot{Z}(\tau)$ of a point-charge of world-line $Z(\tau)$, with $\tau$ the proper-time, and by replacing the radial distance $r$ by the retarded distance $\xi$.  Then
\begin{equation} \label{gen:6}
  \phi^{\text{E}}(\tau,\xi) \DEF \frac{e}{\xi} \dot{Z}(\tau)
                      \UPS(\xi) \Bigr|_{\tau=\tau_r},
\end{equation}
is a proper distributional representation of the Li\'enard-Wiechert potential while
\begin{equation} \label{gen:7}
  \Phi^{\text{E}}(\tau,\xi) = \frac{e}{\xi} \log(\xi/a)\dot{Z}(\tau)
                   \UPS(\xi) \Bigr|_{\tau=\tau_r},
\end{equation}
is a generating function from which it derives distributionally.  Both $\phi^{\text{E}}(\tau,\xi)$ and $\Phi^{\text{E}}(\tau,\xi)$ are four-vectors, $\partial_\xi$ is an invariant scalar operator, and $\tau_r$ the retarded time.

  From a physical point of view, however, the above defined generating functions $\Phi$, as well as the differential relationship between $\Phi$ and $\phi$, have no physical meaning.  In particular, the logarithm $\log(\xi/a)$ where $a$ is an invariant constant does not relate to any simple physical or geometrical quantity, and the partial derivative $\partial_\xi$ has no four-dimensional interpretation.   Therefore, in view of the mathematically fundamental character of Schwartz's structure theorem, the question arises whether there could exist a generating function $\Phi \in \mathcal{C}^0(\mathbb{R}^4)$ and a differential operator $\mathbf{D}$ such that both $\Phi(\tau,\xi)$ and the relation $\phi(\tau,\xi) = \mathbf{D} \Phi(\tau,\xi)$ have physically simple, natural, and relativistically invariant four-dimensional meanings.  That this is indeed the case is quite remakable, and will now be shown.

\section{Definitions}
\label{def:0} \setcounter{equation}{0}

Let $Z_\MU(\tau)$ be the relativistic four-position of an arbitrarily moving point in space-time, and $X_\MU$ a point of observation.  These two points are said to be causally related if the interval $R_\MU$ between $X_\MU$ and $Z_\MU$ is a null four-vector, i.e., iff
\begin{eqnarray}\label{def:1}
    R_\MU \DEF X_\MU - Z_\MU,
   \qquad\mbox{is such that}\qquad
    R_\MU R^\MU = 0.
\end{eqnarray}
Let us now consider the successive proper-time derivatives of $Z_\MU(\tau)$, i.e., the four-velocity $\dot{Z_\MU}$, four-acceleration  $\ddot{Z_\MU}$, and four-biacceleration $\dddot{Z_\MU}$ of the charge, to which three invariants are associated: $\dot{Z}_\MU R^\MU, \ddot{Z}_\MU R^\MU$, and $\dddot{Z}_\MU R^\MU$.  The first one is called the retarded distance,
\begin{eqnarray}\label{def:2}
          \xi \DEF   \dot{Z}_\MU R^\MU,
\end{eqnarray}
which enables to introduce a `unit' null four-vector $K_\MU$ defined as
\begin{eqnarray}\label{def:3}
          K_\MU(\tau,\vartheta,\varphi) \DEF R_\MU/\xi,
\end{eqnarray}
and the so-called acceleration invariant
\begin{eqnarray}\label{def:4}
      \kappa \DEF  \ddot{Z}_\MU K^\MU.
\end{eqnarray}

In order that any effect observed at the position $X_\MU$ is causally related to $Z_\MU(\tau)$, all partial derivatives must be calculated under the condition $R_\MU R^\MU = 0$, which insures causality.  For an expression $E = E(X_\MU,\tau)$, where the argument $X_\MU$ corresponds to an explicit dependence on $X_\MU$, and $\tau$ to the proper time, this condition leads to the covariant differentiation rule
\begin{eqnarray}\label{def:5}
   \partial_\MU E(X_\MU,\tau) = \partial_\MU E(X_\MU)
                              + K_\MU \dot{E}(\tau) \Bigr|_{\tau=\tau_r}. 
\end{eqnarray}
In this equation the condition  $\tau=\tau_r$ implies that all quantities are evaluated at the retarded proper time $\tau_r$.  In the following, for simplicity,  this condition will be specified explicitly only for the main equations.

  To physically interpret the results which will be derived in the following, we need {\it explicit} representations of all retarded quantities, and in particular of $\xi$, $K_\MU$ and $R_\MU$.  For $\xi$ that is easy because in three-dimensional notation the retarded distance \eqref{def:2} reads
\begin{eqnarray}\label{def:6}
          \xi = |\vec{x} - \vec{z}\,| \gamma(1-\vec{\rho}\cdot\vec{\beta}\,),
\end{eqnarray}
where $\vec{\rho}$ is the unit vector in the direction of $\vec{x} - \vec{z}$. The retarded distance is therefore proportional to an absolute value, and for that reason has a discontinuous derivative when $\vec{x} \rightarrow \vec{z}$, i.e., at $\xi=0$ where a second partial differentiation leads to a $\delta$-function.

   For $K_\MU$, and thus $R_\MU = \xi K_\MU$, which are null four-vectors, explicit representations are unavailable in tensor formalism.   Explicit representations of null-quantities involve matrices (e.g., Pauli or Dirac matrices), Clifford number, or, more efficiently, biquaternions \cite{WEISS1941-}.  In our case we only need the biquaternion representations of the null four-vectors $K$ and $R=\xi K$.  For instance,
\begin{eqnarray}\label{def:7}
      X - Z = R =  \xi \mathcal{L} (i + \vec{\nu}\,) \mathcal{L}^+.
\end{eqnarray}
Here $\vec{\nu} = \vec{\nu}(\vartheta,\varphi)$ is a unit vector so that $i + \vec{\nu}$, and consequently $R$ and $K$, are null biquaternions. The biquaternion $\mathcal{L}=\mathcal{B}\mathcal{R}$ corresponds to a Lorentz transformation with boost $\mathcal{B}$ and rotation $\mathcal{R}$ such that $\mathcal{L}\CON{\mathcal{L}}=1$  and  $\mathcal{L}\mathcal{L}^+=\mathcal{B}\mathcal{B}^+=\dot{\mathcal{Z}}$. 

   Equation~\eqref{def:7} is therefore an explicit parametrization of $R$ in terms of four variables: The invariants $\xi$ and $\tau$, and the two angles $\vartheta$ and $\phi$ characterizing the unit vector $\vec{\nu}$ in the instantaneous rest frame of the point $Z$.  Thus, in that frame, i.e., when $\mathcal{L}=1$, the interval $X-Z$ reduces to the quaternion $\xi (i + \vec{\nu}\,)$, which shows that in that frame the vector $\xi \vec{\nu}$ corresponds to the ordinary radius vector $\vec x - \vec z$, and that the distance $\xi$ appears to an observer at rest in that frame as the ordinary distance $|\vec x - \vec z|$.  On the other hand, when $\mathcal{L}\neq 1$, Eq.~\eqref{def:7} provides a general parametrization of the null-interval $R$, that is, in geometrical language, of the `light-' or null-cone originating from $Z$.

    A remarkable property of \eqref{def:7}, which was first noted by Paul Weiss \cite{WEISS1941-}, is that arbitrariness in the choice of the origins of the angles $\vartheta$ and $\varphi$ corresponds to what he called an `internal rotation,' i.e., an invariance with respect to the group $SO(3),$ which enables to write
\begin{eqnarray}\label{def:8}
      i + \vec{\nu}
   =  \mathcal{W} (i + \vec{\nu}_0) \CON{\mathcal{W}},
\end{eqnarray}
where $\mathcal{W}$ such that $\mathcal{W}\CON{\mathcal{W}}=1$ is a unit real quaternion, i.e., an element of $SU(2)$, and $\vec{\nu}_0$ a fixed unit vector.

 Finally, for convenience, it is useful to write the null biquaternion $i + \vec{\nu}$ in terms of an idempotent $\sigma$ such that $\sigma \sigma = \sigma$ and $\sigma \CON{\sigma} = 0.$  Therefore, we write
\begin{eqnarray}\label{def:9}
      i + \vec{\nu} = 2i \CON{\sigma},
      \qquad\mbox{where}\qquad
      \sigma \DEF \frac{1}{2}(1 + i \nu),
\end{eqnarray}
so that, when $\mathcal{R}=1$,
\begin{eqnarray}\label{def:10}
         K(\tau,\vartheta,\varphi) = 2i \, \mathcal{B}(\tau) \,
        \CON{\sigma}(\vartheta,\varphi) \, \mathcal{B}^+(\tau).
\end{eqnarray}

\section{The causal interval $R$ as a generating function}
\label{cau:0} \setcounter{equation}{0}

Let us consider as generating function the  $\mathcal{C}^0(\mathbb{R}^4)$ function
\begin{eqnarray}\label{cau:1}
      \Phi_\MU(\tau,\xi,\vartheta,\varphi) \DEF 
 \frac{1}{2} e R_\MU(\tau,\xi,\vartheta,\varphi) \UPS(\xi)\Bigr|_{\tau=\tau_r},
\end{eqnarray}
where, consistent with \eqref{def:6}, $\UPS$ takes into account the discontinuity of the spatial derivatives of $\xi$ at $\xi=0$. Then, with $\square \DEF \partial^\MU \partial_\MU$ defined as the causal d'Alembertian operator obtained by iterating \eqref{def:5}, we obtain after a number of lines of calculations that are most easily done using the quaternion methods exposed in \cite{GSPON2006D}, 
\begin{equation} \label{cau:2}
  \square \Phi_\MU = e\frac{1}{\xi}\dot{Z}_\MU \UPS(\xi)
                   + e\frac{1}{2}K_\MU \Bigl( (4-6\xi\kappa)\UPS'(\xi)
                   + (1-2\xi\kappa)\xi\UPS''(\xi) \Bigr),
\end{equation}
where everything is evaluated at $\tau=\tau_r$.

  The first term of \eqref{cau:2} is evidently the Li\'enard-Wiechert potential \eqref{gen:6}, a result which by itself is not new because the relation $\square R = 2/\xi$ has been derived by many physicist and mathematicians as a byproduct of calculations using retarded coordinates, although this identity seems never to have been interpreted in the context of distributions and nonlinear generalized functions theories as we do here.

   The second term in \eqref{cau:2} is of course the more interesting one.  Let us first discuss it in the context of Schwartz distribution theory.  As $\UPS' = \delta$ by \eqref{gen:4}, we have then $\UPS'' = - \delta/\xi$, which enables to rewrite \eqref{cau:2} as
\begin{equation} \label{cau:3}
   \square \Phi_\MU = \phi^{\text{E}}_\MU(\tau,\xi)
                    + \phi^{\text{K}}_\MU(\tau,\xi,\vartheta,\varphi),
\end{equation}
where $\phi^{\text{E}}_\MU(\tau,\xi)$ is given by \eqref{gen:6}, and
\begin{equation} \label{cau:4}
   \phi^{\text{K}}_\MU(\tau,\xi,\vartheta,\varphi)
            = e\frac{(3-4\xi\kappa)}{2} K_\MU \delta(\xi) \Bigr|_{\tau=\tau_r}.
\end{equation}
Clearly, as the null four-vector $K_\MU$ is a finite function of the angular variables, evaluating  \eqref{cau:3} on a test function $T \in \mathcal{D}(\mathbb{R}^3)$ implies that the first term is the equivalent to the Li\'enard-Wiechert potential \eqref{gen:6}, and that the second term is zero.  Consequently, in  Schwartz distribution theory, all terms beyond the first one in \eqref{cau:2} and \eqref{cau:3} can be ignored, i.e., symbolically
\begin{equation} \label{cau:5}
  \langle \phi^{\text{K}}_\MU | T \rangle = 0.
\end{equation}

  If we now interpret \eqref{cau:3} as a nonlinear generalized function $\square \Phi_\MU$ will still be a distribution but the second term, i.e., \eqref{cau:4}, will no more be ignored.  Indeed, in the nonlinear generalized function context we consider the possibility that in some physical theory that second term is nonlinearly operated on in such a way that it yields a non-zero contribution.  Thus we consider generalized test functions of the form $F(\xi) = T(\xi)/\xi^n$, where $T \in \mathcal{D}(\mathbb{R}^3)$, such that $\phi^{\text{K}}_\MU$ evaluated on $F$ gives a finite non-zero result.  Evidently, as the $\xi$ integration element is $\xi^2d\xi$ in $\mathbb{R}^3$, this happens when $n=2$ so that
\begin{equation} \label{cau:6}
  \langle \phi^{\text{K}}_\MU | \frac{1}{\xi^2} T \rangle
   = \langle \frac{1}{\xi^2} \phi^{\text{K}}_\MU |  T \rangle
   \neq 0.
\end{equation}
Consequently, we are led to seek a physical interpretation for the contribution $\xi^{-2} \phi^{\text{K}}_\MU(\tau,\xi,\vartheta,\varphi)$, what will be done now.

\section{Electroweak unification}
\label{ele:0}

  Referring to the decomposition \eqref{cau:3} we have found that the first term in $\square \Phi_\MU$, i.e., in the d'Alembertian of the causal interval $R=X-Z$  multiplied by $e/2$, corresponds to the electromagnetic four-potential at $X$ of an arbitrarily moving point-charge of world-line $Z(\tau)$.  On the other hand, the second term corresponds to the four-potential of a different interaction which is nevertheless necessarily associated to electromagnetic interactions, and which produces a finite effect provided $\square \Phi_\MU$ is interpreted as a nonlinear generalized function.

  In order to identify that non-electromagnetic interaction, and to express it in a form directly comparable to $\phi^{\text{E}}_\MU(\tau,\xi)$, i.e.,\footnote{For simplicity we do not specify any more that everything is evaluated at $\tau=\tau_r$.}
\begin{equation} \label{ele:1}
  \phi^{\text{E}}_\MU(\tau,\xi) = e\frac{1}{\xi} \dot{Z}_\MU(\tau) \UPS(\xi),
\end{equation}
we define the potential
\begin{equation} \label{ele:2}
   \phi^{\text{W}}_\MU(\tau,\xi,\vartheta,\varphi)
   = e\frac{3}{2}\frac{\lambda^2}{\xi^2} K_\MU \delta(\xi),
\end{equation}
which is obtained by multiplying \eqref{cau:4} by the factor $(\lambda/\xi)^2$, where the constant $\lambda$ is an invariant length, and that we mathematically interpret as a Schwartz distribution, so that (in particular) the factor $\xi\kappa$ in \eqref{cau:4} could be ignored.   Thus, for an arbitrarily moving point $Z(\tau)$, the potentials $\phi^{\text{E}}$ and $\phi^{\text{W}}$ are distributions that do not depend on any proper time derivative of $Z$ beyond $\dot{Z}$.

    To physically interpret $\phi^{\text{E}}$ and $\phi^{\text{W}}$ we use the Fourier transform formulas
\begin{align}
\label{ele:3}
   \mathsf{F}\Bigl(\frac{1}{\xi} \Bigr) &= 
   \iiint \frac{d^3\xi}{4\pi} \exp(i\vec{q}\cdot\vec{\xi}\,) \frac{1}{\xi}
   = \frac{1}{q^2},  \\
\label{ele:4}
   \mathsf{F}\Bigl(\frac{\delta(\xi)}{\xi^2} \Bigr) &=
   \iiint \frac{d^3\xi}{4\pi} \exp(i\vec{q}\cdot\vec{\xi}\,)
   \frac{\delta(\xi)}{\xi^2} = 1.
\end{align}
to calculate
\begin{align}
\label{ele:5}
  \Pi^{\text{E}}_\MU &\DEF \mathsf{F}\Bigl(e \phi^{\text{E}}_\MU \Bigr)  
   =  \frac{e^2}{q^2} \dot{Z}_\MU,  \\
\label{ele:6}
  \Pi^{\text{W}}_\MU &\DEF \mathsf{F}\Bigl(e \phi^{\text{W}}_\MU \Bigr)
   =  \frac{3}{2} e^2\lambda^2 K_\MU.
\end{align}
Then, if $q=|\vec{q}\,|$ is the modulus of the transferred momentum, $\Pi^{\text{E}}_\MU$ is of course the well-known `propagator' of the electromagnetic field, as will shortly be further confirmed.

  As for $\Pi^{\text{W}}_\MU$, which is just a constant multiplying the null-vector $K_\MU$, we first remark that if we take for $\lambda$ the Compton-length $\hbar c/M_{\text{W}}$ of a particle of mass $M_{\text{W}}$, that constant takes the form
\begin{equation}\label{ele:7}
                 \frac{3}{2} \frac{e^2}{\hbar c}
                 \frac{(\hbar c)^3}{M_{\text{W}}^2}
                 \approx \frac{G_{\text{F}}}{\sqrt{2}},
\end{equation}
which within a factor of order unity is identical to that of the Fermi constant of weak interactions provided $M_{\text{W}}$ is the mass of an intermediate vector boson.  Consequently, we see that $\Pi^{\text{W}}_\MU$ could correspond to weak interactions, which is plausible because in the Fermi limit (i.e., at low-energies) the propagator of that interaction is a constant which does not depend on the transferred momentum $\vec{q}$.

  To confirm that $\Pi^{\text{W}}_\MU$ does indeed correspond to the propagator of weak interactions in the Fermi limit, we rewrite (\ref{ele:5}--\ref{ele:6}) in biquaternion formalism, i.e., setting $G_{\text{W}} = 3 e^2/M_{\text{W}}^2$,
\begin{align}
\label{ele:8}
  \Pi^{\text{E}} &=  i\frac{e^2}{q^2} \dot{Z}
                  =  i\frac{e^2}{q^2} \mathcal{B}\mathcal{B}^+,  \\
\label{ele:9}
  \Pi^{\text{W}} &=  i \frac{G_{\text{W}}}{2} K
                  =  iG_{\text{W}} \mathcal{B}\CON{\sigma}\mathcal{B}^+,
\end{align}
were we used the identity $\dot{Z} = \mathcal{B}\mathcal{B}^+$ and equation \eqref{def:10}. Then, if $D_{\text{i}}$ and $D_{\text{f}}$ are two biquaternions characterizing some initial and final states, the transition amplitude between them is given by the invariant scalar\footnote{This is a rigorous but simplified presentation: here the $D$ fields are not actual solutions of the Lanczos or Dirac equations, but biquaternion factors containing their spin and isospin contents.}
\begin{equation}\label{ele:10}
  T_{\text{fi}} = i \mathbb{S} [ D_{\text{f}}^+ \CON{\Pi} D_{\text{i}} ],
\end{equation}
in which for simplicity we omitted the energy-conserving $\delta$-function, etc.

  In the case of $\Pi^{\text{E}}$, using Lorentz invariance to set $\mathcal{B}=1$, we get
\begin{equation}\label{ele:11}
  T_{\text{fi}}^{\text{E}} = - \frac{e^2}{p^2}
                               \mathbb{S} [ D_{\text{f}}^+ D_{\text{i}} ],
\end{equation}
which as expected corresponds to the electromagnetic transition amplitude in first order perturbation theory. 

   In the case of $\Pi^{\text{W}}$, in which the null quaternion $\sigma$ appears, we need first to recall the theorem stating that any biquaternion $B$ can uniquely be written as $B = Q_1 \sigma + Q_2 \CON{\sigma}$, or $B = \sigma Q_3  + \CON{\sigma} Q_4$, where $Q_1$ to $Q_4$ are real quaternions, and where $\sigma$ is a given idempotent, i.e., a parameter  \cite{GSPON1993B}.  Ignoring for the moment that $\vec{\nu}$ in $\sigma$ defined by \eqref{def:9} is not fixed but in the spatial direction of the interval $X-Z$, we can write
\begin{equation}
\label{ele:12}
   D_{\text{f}}^+ = \CON{Q}_{\text{f,L}}      \sigma
                  + \CON{Q}_{\text{f,R}} \CON{\sigma},
      \qquad\mbox{and}\qquad
   D_{\text{i}} =      \sigma  Q_{\text{i,L}}
                + \CON{\sigma} Q_{\text{i,R}}.
\end{equation}
where the labels L and R conventionally refer to left- and right-handness so that $D_{\text{i}} = D_{\text{i,L}} + D_{\text{i,R}}$ and $D_{\text{f}}^+ = D_{\text{f,L}}^+ + D_{\text{f,R}}^+$.  Then, inserting $\Pi^{\text{W}}$ in \eqref{ele:10}, and setting $\mathcal{B}=1$ again, we get 
\begin{equation}\label{ele:13}
  T_{\text{fi}}^{\text{W}} = - G_{\text{W}}
                              \mathbb{S} [ D_{\text{f}}^+ \sigma D_{\text{i}} ]
= - G_{\text{W}} \mathbb{S} [ D_{\text{f,L}}^+ D_{\text{i,L}} ],
\end{equation}
which implies that \emph{only the left-handed parts} of the fields $D_{\text{i}}$ and $D_{\text{f}}$ are present in the transition amplitude, a feature characteristic of weak interactions.  Thus, if the idempotent $\sigma=\tfrac{1}{2}(1 + i\vec{\nu}\,)$ were \emph{fixed}, just like the chirality idempotent $\tfrac{1}{2}(1 + \gamma_5)$ is a fixed matrix in the customary formulation of Dirac's theory, the propagator $\Pi^{\text{W}}$ would correspond to weak interactions.

   Fortunately, this is indeed possible:  it suffice for this, as the point $Z(\tau)$ moves relative to $X$, that an internal Weiss rotation \eqref{def:8} is made so that at any time
\begin{equation}\label{ele:14}
    \sigma(\tau) = \mathcal{W}(\tau) \sigma_0 \CON{\mathcal{W}}(\tau),
      \qquad\mbox{where}\qquad 
    \sigma_0 = \tfrac{1}{2}(i + \vec{\nu}_0).
\end{equation}
However, replacing $\sigma$ by $\sigma(\tau)$ in \eqref{ele:13}, the transition amplitude is no more invariant!  But, again, there is a simple solution: invariance is restored if at the same as the $SO(3)$ transformation \eqref{ele:14} is made on $\sigma$, we postulate that a compensating $SU(2)$ transformation is made on the $D_{\text{i}}$ and $D_{\text{f}}$ fields, i.e., if
\begin{equation}\label{ele:15}
   D_{\text{f}}^+(\tau) =  D_{\text{f,0}}^+ \CON{\mathcal{W}}(\tau),
      \qquad\mbox{and}\qquad
   D_{\text{i}}(\tau) =  \mathcal{W}(\tau) D_{\text{i},0}.
\end{equation}
Thus, instead of \eqref{ele:13}, we finally get
\begin{equation}\label{ele:16}
  T_{\text{fi}}^{\text{W}}
 = - G_{\text{W}} \mathbb{S} [ D_{\text{f,0}}^+ \sigma_0 D_{\text{i,0}} ]
= - G_{\text{W}} \mathbb{S} [ D_{\text{f,0,L}}^+ D_{\text{i,0,L}} ],
\end{equation}
where  $D_{\text{i,0}}$ and $D_{\text{f,0}}$ correspond to solutions of the Lanczos or Dirac equations parameterized in terms of the idempotents $\sigma_0$ and  $\CON{\sigma}_0$ which are now constant and invariant, because the arbitrary unit vector $\vec{\nu}_0$ is fixed.

    In the language of the `standard model,' the property that the equations of motion defining the world-line $Z(\tau)$ are such that the fields $D_{\text{i,0}}$ and $D_{\text{f,0}}$ are invariant under the $SU(2)$ transformations \eqref{ele:15} is called `non-Abelian gauge invariance,'\footnote{That invariance requires of course the introduction of gauge fields to compensate locally for the $\tau$ dependence of $\mathcal{W}(\tau)$.  But we do not need to consider these implications for our present purpose.}  Here, this invariance was used to fix the vector $\vec{\nu}_0$, which therefore corresponds to our choice of the gauge.  We have therefore encountered what constitutes the final essential feature of weak interaction that was needed for our interpretation to be plausible, although that feature did not derive form our considerations on generalized function theory, but had to be postulated in order to obtain the fully invariant transition amplitude \eqref{ele:16}.  Finally, we can also verify that $G_{\text{W}}$ gives a reasonable numerical estimate of $M_{\text{W}}$ by solving
\begin{equation}\label{ele:17}
  G_{\text{W}} = 3 \frac{e^2}{\hbar c}
                 \frac{(\hbar c)^3}{M_{\text{W}}^2}
               = \frac{G_{\text{F}}}{\sqrt{2}},
\end{equation}
where $G_{\text{F}}/(\hbar c)^3 = 1.166 \times 10^{-5} \text{ GeV}^{-2}$ gives $M_{\text{W}} = 51$ GeV to be compared to the measures values of $80$ and $91$ GeV for the $W^\pm$ and $Z_0$ bosons.

\section{Discussion and outlook}
\label{dis:0}

The purpose of this paper was not to present a `complete model' of electroweak interactions that could be directly compared to experiment.  Rather, our intent was to show that electroweak unification can be explained by relating the interactions (i.e., here, non-gravitational forces) between two space-time points $X$ and $Z$ to their causal interval $R=X-Z$, which is a $\mathcal{C}^0(\mathbb{R}^4)$ function, so that it can be taken as the generating function of a sequence of distributions obtained by successive causal differentiations.  It then turned out that the causal d'Alembertian of that generating function directly led to the potential of electromagnetic interactions, to which the potential of another interaction was necessarily added to.

Interpreting that other interaction and showing that it corresponds to weak interactions --- at least in the limit in which the Fermi model of weak interactions is valid  ---  was the main task performed in the paper.  That interpretation required that this non-electromagnetic interaction be interpreted in the context of nonlinear generalized functions.  That meant that terms that would be classified as `negligible' in distribution theory had actually to be retained, a process that appears to lead to new results and to provide a deeper understanding of many problems ranging from classical electrodynamics to quantum field theory \cite{GSPON2008B, GSPON2006C, COLOM2008-}.

    In order to go from the `principle' given in the introduction to a `complete model' of electroweak interactions much works remains to be done.  But if at the same time as nonlinear generalized functions are introduced, a formalism in which isospin is included right from the beginning is used, such as in Lanczos's generalization of Dirac's equation \cite{GSPON1993B,GSPON1998B,GSPON2001-}, the task could be not as daunting as it may seem.\footnote{The ideas presented in this paper may have significant implications for other type of interactions, including gravitational ones, in which the electric-like `linear distributional forces' may as well have to be supplemented by weak-like `nonlinear generalized forces...'}

\section{References}
\label{biblio}

\end{document}